\documentclass[12pt]{article}
\usepackage{amsmath,amsfonts}
\usepackage[numbers,sort&compress]{natbib}
\usepackage{times}
\usepackage[left=2.54cm,top=2.54cm,right=2.54cm,bottom=2.54cm,bindingoffset=0.0cm]{geometry}
\usepackage{setspace}
\usepackage{mdframed}

\usepackage{graphicx}
\usepackage{amssymb}
\usepackage{vaucanson-g}
\usepackage{lmodern}
\usepackage{amsmath}
\usepackage{amsthm}
\usepackage{amsfonts}
\usepackage{epstopdf}
\usepackage{tikz}

\usepackage{hyperref}

\title{Common Knowledge on Networks}
\author{Torrin M. Liddell\footnote{Department of Psychological and Brain Sciences, Indiana University, 1101 E.\ 10th St., Bloomington IN 47405. {\tt torrin.liddell@gmail.com}} \and Simon DeDeo\footnote{School of Informatics and Computing, Indiana University, 901 E. 10th St., Bloomington, IN 47408 \& Santa Fe Institute, 1399 Hyde Park Road, Santa Fe, NM 87501. {\tt sdedeo@indiana.edu}}}
\date{\today}

\begin{document}

\maketitle
\begin{abstract}
\noindent
Common knowledge of intentions is crucial to basic social tasks ranging from cooperative hunting to oligopoly collusion, riots, revolutions, and the evolution of social norms and human culture. Yet little is known about how common knowledge leaves a trace on the dynamics of a social network. Here we show how an individual's network properties---primarily local clustering and betweenness centrality---provide strong signals of the ability to successfully participate in common knowledge tasks. These signals are distinct from those expected when practices are contagious, or when people use less-sophisticated heuristics that do not yield true coordination. This makes it possible to infer decision rules from observation. We also find that tasks that require common knowledge can yield significant inequalities in success, in contrast to the relative equality that results when practices spread by contagion alone.
\end{abstract}

\noindent
In the game known as Rousseau's Stag Hunt~\citep{Skyrms2001}, participants must decide between hunting a stag, or rabbits, before knowing what the others have chosen to do. Rabbits can be caught by individuals, but provide a lower reward than a stag. A stag can be caught, but only if others also decide to join the hunt, and a failed hunt is a serious setback. 

Even when an individual desires to join a hunt, she may not know that others know this. In the face of this uncertainty, she may go after rabbits instead and gain a meager, but relatively assured, benefit. Yet the other members of her group followed similar logic. All preferred to hunt stag, but all ended up hunting rabbits. This is the problem of common knowledge.


The problem of the stag hunt is widespread in the modern world. An entrepreneur decides between a stable salary at a major corporation, or joining with others in a San Francisco startup. In determining whether potential team-mates are willing to forego their own back-up plans and commit to joint effort, he faces the same problem as Rousseau's hunter. So, too, does an individual considering joining a demonstration against an unpopular regime, where small crowds may be violently dispersed, but safety, and political change, lies in numbers~\cite{Chwe1999}.

Much work has been done on the mechanisms that can drive cooperation, but ensuring that all individuals involved have the capability to coordinate is an equally difficult problem~\citep{Bowles2011}. In the psychological sciences, this is often studied in the context of hidden profile problems, where optimal decision-making requires the synthesis of both public and private information from multiple individuals. Successful completion requires the generation of common knowledge of the private information, a task which proves to be very difficult~\citep{Gigone1993}. However, under some circumstances, individuals are capable of surmounting this problem and building the proper common knowledge to reach the correct conclusion~\citep{Stasser1992,Reimer2010}.
%
%

%

Here, we study how this bounded logic plays out in real-world scenarios, with three increasingly sophisticated heuristics for solving the common knowledge problem in a social network. We provide a brief introduction to the theory of common knowledge. We then specify a series of three decision rules that interpolate between simple copying of others (contagion), and full common knowledge. We simulate these decision rules on realistic social networks. This enables us to derive a behavioral signature that indicates that individuals are making use of common knowledge in their decision-making process. We analyse both group-level outcomes, and the network properties of individuals that predict task participation.

When common knowledge is required, we find that an individual's participation in a task is predicted by high levels of clustering in their local neighborhood, and being in a central location in the graph. Importantly, this behavioral signature is not present in other potential network models for the adoption of behaviors. We thus provide a new diagnostic, at the network level, for tasks that require common knowledge. We also find that these more sophisticated rules lead to unequal outcomes: only some individuals are consistently able to join in risky cooperative tasks.

\section{The Theory of Common Knowledge}

At the heart of the stag hunt is a problem concerning the minds of others. If a group of three is needed, it is not enough for me to know that $i$ and $j$ are willing to join; I will want to know that $i$ knows that $j$ is willing---otherwise, I will worry that, presuming the absence of $j$, she will fail to appear. And, of course, if I want to know this, it is natural to assume that $i$ will want parallel knowledge: that $j$ knows that I am willing. And if I have reason to doubt that $i$ knows that $j$ knows that I am willing, again---given sufficiently low risk-tolerance---the hunt will fail. 

I may inform $i$ that $j$ knows that I am willing; but, of course, now I must inform $j$ that $i$ knows this, and return to $i$ to inform her that $j$ knows that---the regress, at least in a formal sense, is infinite. What is needed is not only for all members of the group to have knowledge of the group's willingness, but also for all members to know that all members know this, and also that all members know that all members know this, and so on. 

This is the \emph{problem of common knowledge}. In order for a group to build common knowledge, all members of the group must communicate with one another, and this communication must be known to all members of the group. Formally, any game with multiple Nash equilibria must solve the common knowledge problem~\citep{Aumann1995}, and Ref.~\citep{Aumann1999} provided a mathematical framework for describing the ideal case. But ordinary human coordination never achieves such an optimal state~\citep{john1992,Gintis2009}. In the presence of imperfect knowledge and finite understanding, we are forced to rely on heuristics such as those we present here. 

In the case above, for example, I might build common knowledge by collecting members of my group together and having them declare their willingness, face-to-face, in public~\citep{Chwe2013}. Or, relying on bounded-rational decision-making, I might be content (in the three person case, with $i$, and $j$) to inform $i$ and $j$ about my willingness (and learn of theirs), relying on a link between $i$ and $j$ to convey their mutual willingness and mine to each other. 

I  may not know (for example) that $i$ knows that $j$ knows that I am willing, but this high a level of ``level-$k$ thinking''~\citep{Camerer2004} is rarely observed to influence real-world behavior, even in high-stakes cases involving the potential for mid-air passenger jet collision~\citep{Lee2012}. In order to simulate this process we present, in the next section, a set of three increasingly sophisticated heuristics, or decision-rules, for how individuals might decide to participate in a cooperative task.

\section{Network Models of Common Knowledge}
\label{methods}

The first rule is simple contagion, in which behaviors travel through the network only by observation of behavior. Contagion is a basic mechanism for how behaviors and practices might be adopted on a network~\citep{Granovetter1978,Watts2002}. In these models, cultural practices and task engagement amount to the spread of an infectious meme.

The second rule is the egoist rule. When following this rule, individuals anticipate the behavior of others, but without simulating their meta-cognition. Individuals believe that the knowledge they possess is naturally shared by anyone they encounter.

The final rule is that of common knowledge. Here, individuals make decisions in line with the reasoning presented above, simulating the meta-cognitive states of the other players and choosing to participate only when others are guaranteed to as well. This rule is realistically satisfiable using a heuristic that assumes that everyone \emph{else} is an egoist.

In the real world, success---the capture of the stag---is never guaranteed. Instead,  individuals must assess the relative rewards of the stag compared to the risk of a failed hunt due to insufficient numbers. We are thus concerned with how decision makers attempt to meet their own criteria for sufficient numbers using one of these three heuristics. In all three cases, as we shall see, it is possible for a player to decide to undertake the joint task, but to find herself without the desired number of participating neighbors.


\subsection{Common features of all rules}

All three rules have two free parameters. The first is the base threshold for behavioral adoption, $\theta$. This baseline threshold regulates the personal threshold, $\theta_i$, for each node $i$. An individual's $\theta_i$ starts equal to the base threshold $\theta$ but is distinct from it and can change over time. Both $\theta$ and $\theta_i$ are positive integers.

The threshold $\theta_i$ dictates the willingness of individual $i$ to participate in the task. An individual node $i$ desires that there be $\theta_i$ likely-to-participate nodes in order to be willing to participate. When the node $i$ is deciding whether to participate or not, the threshold of $i$ ($\theta_i$) is consulted. The node examines its neighbors, and if at least $\theta_i - 1$ are judged to be likely-to-participate, the node also participates.

The threshold $\theta_i$ has a number of interpretations that, while conceptually distinct, are equivalent for our purposes. In cases where the probability of a reward depends on the number of participants, we can think of $\theta_i$ as a measure of an individual's risk tolerance. In cases where the reward is a step function in the number of participants, then $\theta_i$ is an individual's (risk-adjusted) estimate of that threshold. Note that in situations where an objective outside criterion of a successful outcome (\emph{e.g.}, catching the stag in a stag hunt) exists, the participation of $\theta_i$ individuals does not necessarily guarantee this objective criterion will be met. However, insofar as the individual threshold is reflective of the individual's model of outcome probabilities as well as the utility associated with the outcomes, participating with $\theta_i$ total participants is still a success in decision procedure, if not in outcome.

The notion of ``likely-to-participate'' is what varies across our three candidate models and forms the basis of the decision rule. As we shall see, individuals may be mistaken about others, and may find themselves participating in a group smaller than their desired threshold.

The only other free parameter is the accidental participation rate $\alpha$. All three models allow for a small chance to accidentally participate in a behavior. This means that even when an individual does not believe there are enough others that are likely to participate, there is a chance (equal to $\alpha$) that the individual will participate anyway. Where relevant, the decision rules described below assume that every node knows only the local structure of the graph; a node is aware of who its neighbors communicate with. This rate is a simple way to model the effects of noisy communication and misunderstandings endemic to real-world systems.

\subsection{The contagion rule}

Contagion represents information transfer solely through the direct observation of behavior: ``likely-to-participate'' is defined as ``participated recently''. Each individual is willing to attempt a task if, and only if, they see enough of their neighbors participating in the previous time step (barring ``accidental'' participation). Behaviors start when an early adopting individual decides they are willing to ``join'' a behavior despite no peers participating in the behavior (represented by the small accidental participation chance described above). In our earlier hunt example, this means that individuals will only go on a hunt if they saw a sufficient number of individuals go on a hunt yesterday. When this process is operating, individuals have very little motivation to participate in a behavior unless others have already done so.

In more formal terms, the threshold of each node represents how many neighbors a node needs to see participating in a behavior before they adopt the behavior. Recall that nodes include themselves when comparing the number of likely participators to their own thresholds. For example, a node with a threshold of three needs to see two other individuals participating before they themselves will participate. This decision rule is identical to the decision rule used in the standard linear threshold model commonly used in the literature~\citep{Granovetter1978,Watts2002}.

\subsection{The egoist rule}

The egoist model represents individuals who perform a partial simulation of the decision making behavior of their neighbors. The simulation is only partial: no attempt to model the other player's meta-cognition is made. 

An egoist individual polls the willingness of his neighbors. If some group of size $k$ are all willing to participate if at least $k$ people total participate, and $k$ is greater than $\theta_i$, then the egoist will participate. From a technical standpoint, the reasoning is insufficient: even though he knows that there is a willing group of sufficient size, he has not confirmed that all members of the group have this information as well. The egoist assumes that everyone knows what he knows, and thereby justifies his name. Behaviors of this form are often seen in experiments~\cite{ross1977false,krueger1994truly} and in some scenarios may be adaptive~\cite{Dawes1989}. 

Under such a rule, the egoist can end up in a scenario where he thinks he is certain that enough individuals will join him, but in fact he turns out to be the only one who has done so. To see how, recall our opening example, where $h$ spoke to each potential hunter individually, and found that there was a sufficient number of hunters such that they would all be happy if they all attended. Under the egoist model, this standard of evidence is sufficient and $H$ attends (and thus potentially attends an under-attended hunt, as described earlier).

In formal terms, a node's threshold represents the minimum size of the local group of willing individuals necessary for the node to participate. A given node $i$ with threshold $\theta_i$ will participate if and only if there is some subset of $i$'s neighbors (including $i$ itself) $G$, such that $G$ has size $k \geq \theta_i$ and also that all group members $g$ have thresholds $\theta_g \leq k$ (\emph{i.e.}, there must be a group with a size at least equal to $i$'s threshold that all have thresholds less than or equal to the group size). 

\subsection{The common knowledge rule}

This model is adapted from a model developed by Chwe~\citep{Chwe1999}, for the problem of common knowledge in collective action. Chwe primarily framed his model in terms of riot participation, but we extend this model to multiple classes of different cooperative phenomena, specifically those where (unlike in Chwe's model) opportunities for participation occur many times over the course of a given simulation.

The common knowledge model describes scenarios where individuals not only attempt to predict the future behavior of other players, but do so in ways that model the knowledge of the other players. Under the common knowledge rule, an individual will only be considered likely-to-participate if they are part of a group of sufficient size for which it is common knowledge that all would be happy to participate if all did so together.

In network terms, this means a node will only participate if the node is a member of a clique composed of $k$ members, where all members of the clique have thresholds less than or equal to $k$. A clique is a group of nodes where all nodes have edges to all other nodes. We will refer to a clique of $k$ individuals as a $k$-clique. Intuitively, the clique requirement matches the concept of all individuals in a group being certain that all nodes know both the thresholds of all other nodes, as well as the knowledge state concerning the thresholds of all other nodes; that is, the thresholds of clique members are common knowledge among all the clique members. Formally, a node $i$ will participate if and only if $i$ is a member of a group $G$ such that $G$ is a $k$-clique and any given group member $g$ has a threshold $\theta_g$ such that $\theta_g \leq k$.

While this rule might appear to require an Aumann-like infinite hierarchy~\citep{Aumann1999}, it is equivalent to an individual modeling her neighbors as using the egoist rule and only participating when the number of certain-to-participate egoist neighbors is sufficient to satisfy her threshold. In the taxonomy of Ref.~\citep{Camerer2004}, this is a form of level-one reasoning, with the zero level decision rule being egoist.

\subsection{Summary of rules}

\begin{figure}
\begin{center}
\includegraphics[scale=0.8]{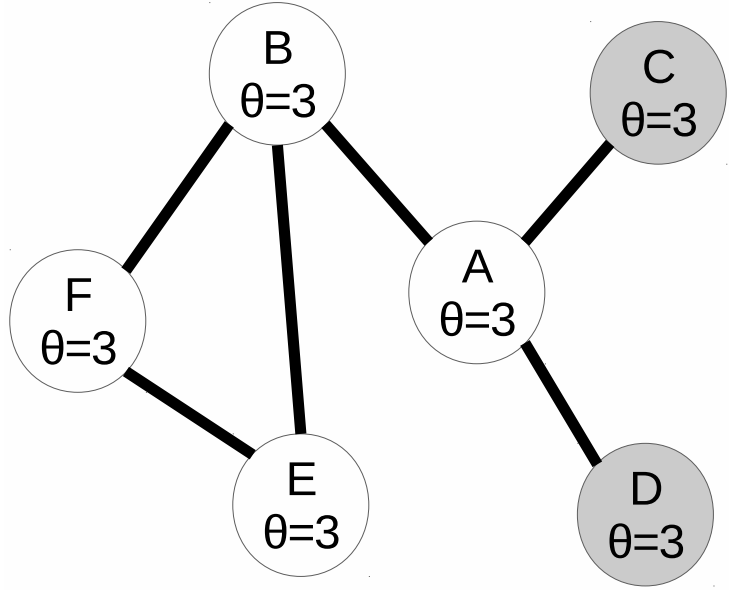}
\end{center}
\caption{A small example network. Capital letters indicate node names, $\theta$ indicates each node's threshold, and filled circles indicate that the node adopted the behavior in the previous time-step. \label{fig:egNet}}
\end{figure}

We review the three decision rules by reference to Figure~\ref{fig:egNet}, which presents a small example network of six nodes, labeled by their thresholds. Filled circles indicate nodes that participated in the previous time step.

First, consider node $A$'s behavior under the contagion rule. $A$ checks if its neighbors $B$, $C$, and $D$ have participated in the previous time step. $C$ and $D$ have done so, and $A$'s threshold for participation is three; $A$ will participate in the next round. Conversely, $C$ has no neighbors who have recently participated, and also has a threshold of 3, and so will not participate in the next round (barring the accidental participation governed by rate $\alpha$).

Now consider the egoist decision rule. $A$ consults the same set of neighbors, but now also considers their thresholds. $A$ has three neighbors, all with threshold three. Including $A$ itself, this makes four individuals willing to participate in a group of three or more, one more than necessary for $A$ to participate, thus $A$, again, in this case, participates. 

Finally, consider $A$ utilizing the common knowledge decision rule. The same set of neighbors are relevant, but now the connections between neighboring nodes are also relevant. $A$ knows there are four individuals willing to participate in a group of three or more, but it also knows that each of these nodes does not have this information, as they do not have edges to one another. Therefore $A$ will not participate. Conversely, node $F$ knows there is a group of three nodes including itself willing to participate in a group of three or more, \emph{and} that this fact is common knowledge among this group. This is because $B$, $E$, and $F$ form a clique of nodes, all whom communicate with one another. $F$ will, under this decision rule, participate.

\subsection{Two Scenarios}
\label{random_graph}

The three rules described above neatly partition the space of decision procedures for the adoption of a behavior or participating in a task. In this section, we place these rules in real-world contexts to see what effects they have on individuals and on the large-scale outcomes of the societies they belong to. 

We consider two distinct scenarios: one, ``the hunting party'', closely mimics the original stag hunt problem. The other, ``adopted fad'', considers cases where individuals might wish to adopt a new cultural practice conditional on the participation of others. These scenarios differ only in what happens to node thresholds after the adoption of the behavior in question. 

In the hunting party case, individuals who participate in a hunt show a subsequent (temporary) disinclination to participate during subsequent time steps. Every time an individual attends a hunt, her threshold is increased by one. If an individual does not attend a hunt and her threshold is above the base threshold, $\theta$, she then decreases her threshold by one. This process stops once the base threshold is reached. This means that participation is usually followed by a decrease in enthusiasm that eventually builds back up to the base threshold. It models an individual who has either sated herself from the gains of successful cooperation, or has become temporarily risk averse subsequent to a failure to reach quorum.

In the case of the adopted fad, adoption of the practice leads to a temporary \emph{increase} in the enthusiasm to continue. When an individual participates in a fad, her threshold immediately drops to the minimum of one. Following this initial drop, her threshold increases by one for each additional round of participation. After an individual stops participation, her threshold decreases by one if it is above the base threshold, or otherwise remains the same. This means that participation is usually followed by further participation, but eventually this participation wanes.

We consider two notions of success. The first is ``participation success'': simple adoption of the behavior, whether or not the individual reaches quorum. Under this notion, the success condition is simply whether or not an outside observer would see an individual attempt to participate in the task at hand (hunting party), or adopt the behavior in question (adopted fad). We can also measure ``quorum success'': the fraction of the time an individual both participates and meets her threshold $\theta_i$. For simplicity, we show here the results for participation success. As we show in the Appendix, all of the major conclusions we reach here also hold for quorum success.

Both the hunting party and adopted fad scenarios share a few basic elements: the structure of the network, the assignment of ``thresholds'', and the rate of random adoption. We consider networks randomly generated using the Watts-Strogatz generation method~\citep{Watts1998}; for our simulations here, each network consists of 100 nodes, with an average of eighteen edges and a $\beta$ (the probability that a local edge becomes a random edge) value of 0.1, implemented in the NetworkX python package~\citep{Hagberg2008}. This generation method produces graph structures more realistically similar to the small-world structure commonly observed in naturally occurring networks~\citep{Watts1998}. All edges are bidirectional; in a cooperative endeavor it is natural to assume two-sided communication.

For the same choice of $\theta$ and $\alpha$, the three decision rules lead to  different overall rates of participation; we thus choose parameter pairs, rule by rule, that lead to the same average rate of participation.

\section{Results}

Given a decision rule, and a scenario, we can study two core topics of interest: (1) the distribution of success within the society as a whole, and (2) the predictors of individual success.

Remarkably, while the hunting party and the adopted fad scenarios seem conceptually distinct, we find very similar results for both society-wide and individual-level outcomes. The main difference is in overall rates of success; because enthusiasm for the hunt declines immediately, it is difficult to get more than 50\% participation among individuals. 

In both scenarios, we find that both common knowledge and egoist decision rules lead to high inequalities in outcome. Societies that rely upon these more cognitively-complex rules have some individuals who are able to achieve the task a large fraction of the time, while others find themselves rarely, if at all, able or willing to participate. Contagion, by contrast, leads to highly equal rates among individuals; not only does the practice reach nearly everyone in the society, it does so at roughly equal rates.

\begin{figure}
\begin{center}
\includegraphics[width=\textwidth]{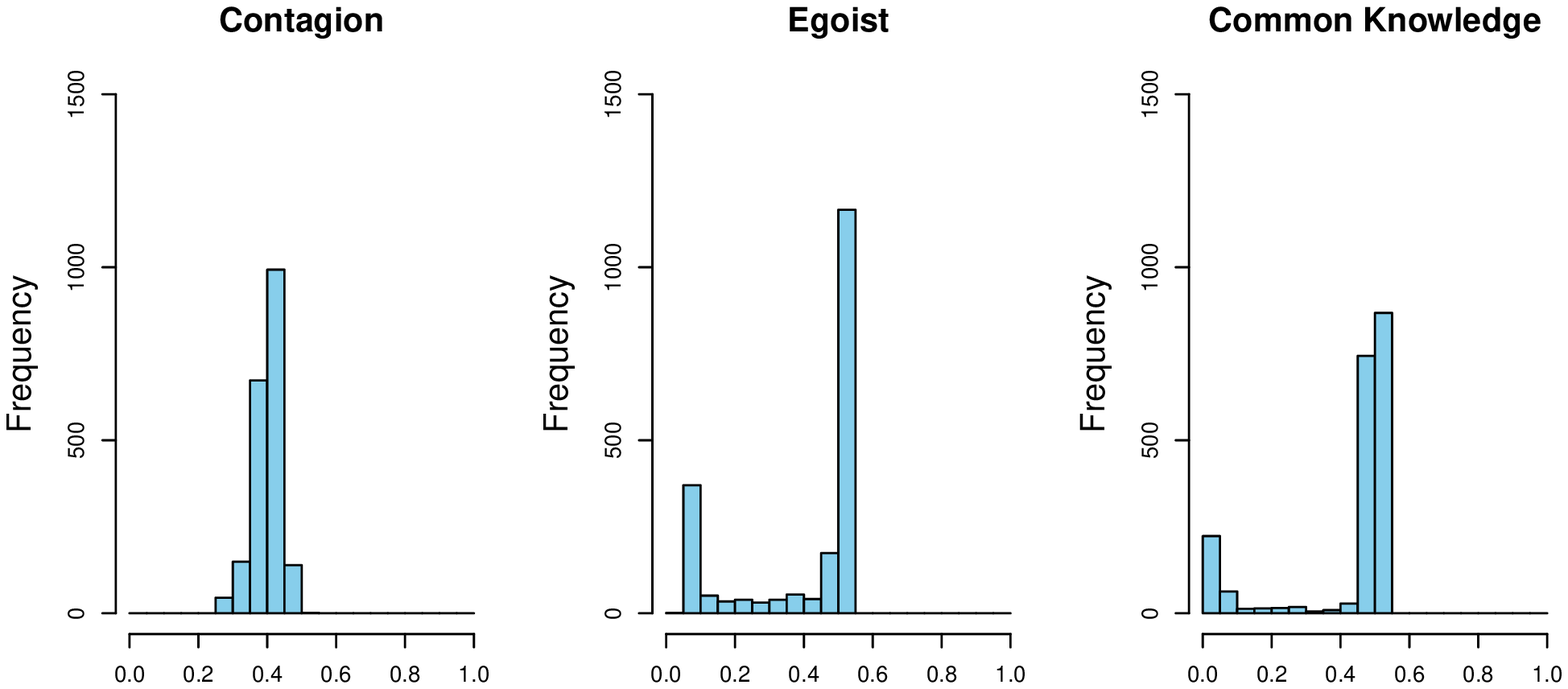} \\
\includegraphics[width=\textwidth]{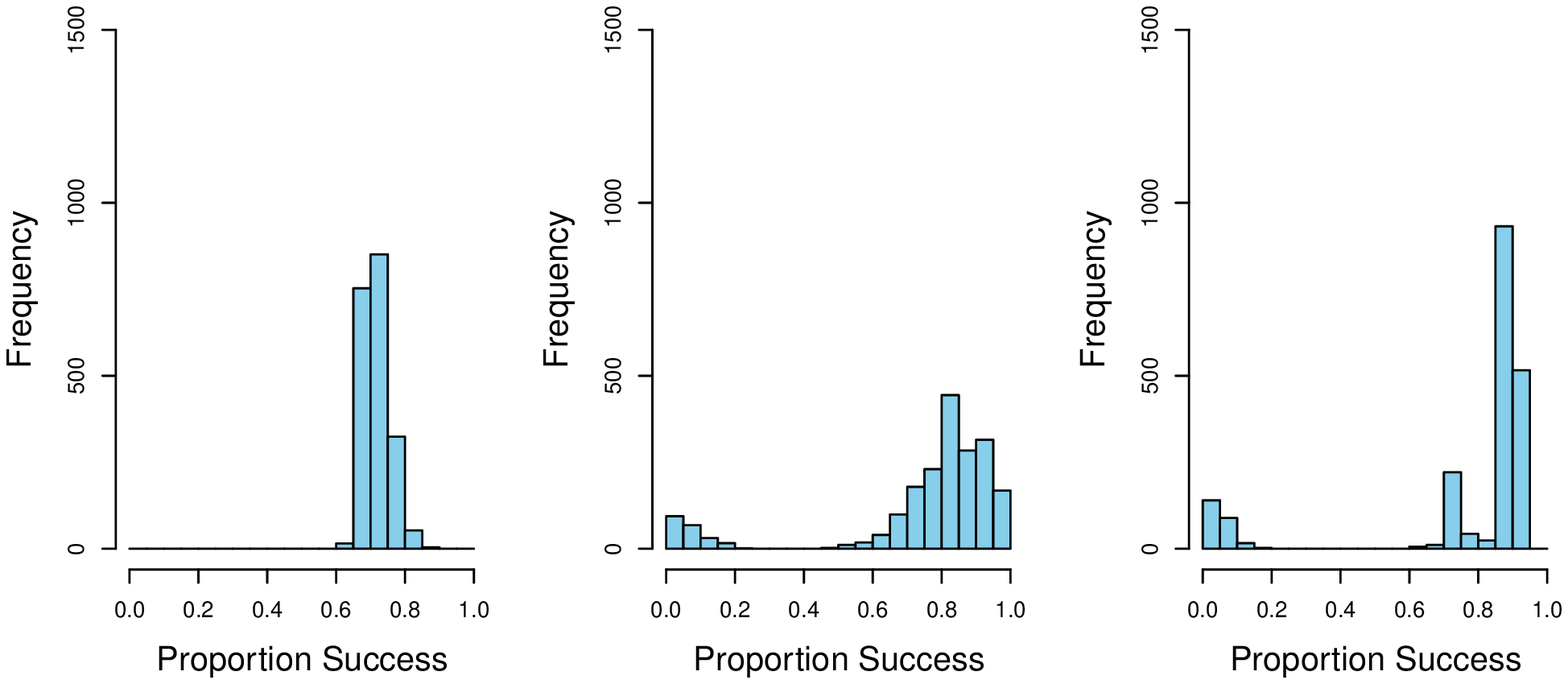}
\end{center}
\caption{\label{fig:Hists} Distributions of individual participation rates for the hunting party scenario (top row) and fad adoption (bottom row). In both cases, egoist and common knowledge decision rules lead to highly unequal outcomes: some nodes have very high rates of participation compared to others, who participate rarely, if at all. Differences in overall averages between the two scenarios represent the greater difficulty of achieving high rates of participation in the hunt scenario, where enthusiasm is sated after a successful hunt.}
\end{figure}

This is shown visually in Fig.~\ref{fig:Hists}; in Table~\ref{gini}, we report the average Gini coefficients~\citep{Gini1912} for these same simulated networks, which confirms these results; the egoist rule shows slightly higher levels of inequality compared to common knowledge and both are much higher than simple contagion. Interestingly, in Fig.~\ref{fig:Hists} a third mode appears in the adopted fad / common knowledge case; these appear to be nodes on the boundary between two densely connected subgroups. These peripheral nodes can be triggered when both subgroups adopt the fad simultaneously.

\begin{table}
\begin{tabular}{l|l|l}
Scenario & Decision Rule & Gini Coefficient \\ \hline
Hunting Party & Contagion & 0.055 \\
& Egoist & 0.221 \\
& Common Knowledge & 0.159 \\ \hline
Adopted Fad & Contagion & 0.029 \\
& Egoist & 0.158 \\
& Common Knowledge & 0.147
\end{tabular}
\caption{\label{gini} Inequalities of outcome: the egoist and common knowledge rules produce far more unequal outcomes (higher Gini coefficients) than the more ``egalitarian'' contagion process.}
\end{table}

Rule-by-rule, differences in participation rate are driven, in large part, by the relative position of individuals in the network. To study this formally, we model each individual node's participation as a linear function of three predictors: 
\begin{enumerate}
\item Degree. The number of neighbors the node has.
\item Betweenness Centrality. The proportion of shortest paths between all node pairs that pass through the node. 
\item Clustering. The proportion of the pairs of neighbors a node has that are themselves neighbors to each other.
\end{enumerate}
We use hierarchical multiple linear regression, with the parameters of the model estimated via Bayesian methods. The model is hierarchical in that we estimate regression parameters for each simulation, with the simulation-level parameters distributed in a higher-level normal distribution.\footnote{For more details on Bayesian multiple regression, see~\citep{Kruschke2015}, Chapter 18. For a more general introduction to Bayesian data analysis see Ref.~\citep{Kruschke2015b}.}

The data (including success and all three predictors) were standardized by $z$-score. This means that the weights are interpretable in terms of standard deviations: a weight of one on a predictor indicates that a one standard deviation increase in the predictor predicts a one standard deviation increase in success. By looking at the weights on each these properties, we can determine which properties matter, and how much.

For the hunting party, degree is a relevant predictor for all decision rules; the higher the degree, the more likely a node is to participate in the task. Betweenness centrality is positively associated with success for the common knowledge rule, negatively related to success in the egoist rule, and unrelated to success in the contagion rule. Clustering is strongly related to success in the common knowledge rule, and is not meaningfully related to success in either of the other two rules. For the adopted fad model, the contagion model is characterized by no credibly non-zero weights, the egoist model is characterized by a strong positive weight on degree, a negative weight on betweenness centrality, and near-zero weight on clustering, and the common knowledge model is characterized by a positive weight on degree, a positive weight on betweenness centrality, and a strong positive weight on clustering. The results of this analysis are shown in Fig.~\ref{props}; details on the Bayesian posterior probabilities are presented in the Appendix.

Despite the structural differences between the hunting party and adopted fad scenarios, a strong weight on clustering remains the primary, and unique, signature of a common knowledge process. To a lesser extent, betweenness centrality is also a unique predictor of common knowledge tasks, again independent of scenario.\footnote{In the egoist model, our results indicate that higher betweenness centrality is predictive of lower levels of success in scenarios similar to the fad case. This suggests that when individuals do not utilize meta-cognitive information about others in deciding to cooperate, being centrally located is worse than being peripheral, holding degree and clustering constant. We suspect that some nodes with high betweenness in this condition are often actually placed between two or more locally connected subgraphs. The spread of participation tends to remain in these smaller subgraphs, meaning that high betweenness individuals have less exposure to this spreading, and so are less likely to participate in a fad.}

\section{Discussion}

Common knowledge may seem like an egalitarian decision rule: to build it requires mutual communication among all members, and when the knowledge is present, all may join in. But this is only a local equality: those who successfully build common knowledge together may be equals, but they have vastly greater ability to succeed compared to those who cannot. A natural distinction between local equality and global inequality emerges from these tightly connected subgroups.

Inequality in human societies has often been linked to multiplicative gains, such as those from return on capital~\cite{piketty2001income,Piketty01082014}, made by a subset of the population. Our results here suggest that network position may play a distinct role in the creation of inequality. Studying network properties  should allow us to make a further distinction between sources of advantage: simple access to others (egoist model), or membership within tightly-knit communities (common knowledge).

Inequality also persists across generations, and success appears to be inherited by a number of distinct mechanisms, including both direct wealth transfer and schooling~\cite{duncan2005apple}. If, as seems likely, social network position is inherited, our results provide another mechanism to explain intergenerational correlations in wealth and status. If I inherit my network position from a previous generation, I will inherit the returns of risky cooperative action.

These results contrast with contagion processes, where those with advantageous positions in the society have less ability to make use of them. Behaviors that spread via contagion on average tend to reach each node with reasonable frequency, with node properties playing only a minor role in overall success.

Local clustering is an important predictor of success in the common knowledge model. It is not just that---as one expects---tightly knit communities enable common knowledge. It is also that related heuristics are insensitive to these same properties. Egotists and copiers do not benefit from being in highly-clustered subgroups.

This relationship can potentially be used to detect the presence of common knowledge tasks. Potential sources of networks include social networking websites, academic co-authorship, records of e-mail or cellphone communication, and sharing of employees between divisions or of residents between cities and nations. Potential tasks of interest may range from literal hunting or decentralized military action to interdisciplinary collaboration, the construction of large-scale open source projects or knowledge commons, and problems of political or social change. The analysis framework presented here allows the detection of common knowledge decision rules in any task that occurs on a network.

The social networks of indigenous populations may provide a particularly clear test of the results presented here. The ``rational ritual''~\cite{Chwe2013} theory provides an explicit account of the connection between group rituals and the desire to generate common knowledge. At the same time, we know from fieldwork that rituals of this form do not necessarily include all members of the society. Only some people get invited to certain parties, rituals, or gatherings. 

In the case of the chicha (beer-drinking) parties among the Tsimane', for example,  invitations are both selective and strongly reciprocal, over and above kin effects~\cite{e15114932}. This leads to an incompletely-connected graph with high local clustering. In societies such as these observation of differential success, as well as more detailed information on correlated timing, could confirm a behavioral process reliant on common knowledge heuristics.

Although we do not consider it here, cognitive load matters. Even in the small village population of Ref.~\cite{e15114932}, it may be difficult for individuals to track the risk-tolerance of all their neighbors. As societies become larger, and the pool of potential cooperators widens, the problem only becomes more acute. The cost of tracking my neighbors, their thresholds, and their connections to each other may lead to errors of judgment. In response to load, individuals may adopt new heuristics not considered here. Individuals may, for example, consider subsets of their neighbors, chosen by an egoist-like rule, and save effort by considering only the links within this subgroup.

Our three rules---contagion, egoist, and common knowledge---interpolate between different levels of cognitive sophistication. Recent work in cultural learning has led to a wide taxonomy of simple heuristics that fall between these three anchor-points, including the copying the decisions of the majority, of those who have previously been successful, or of a random individual~\cite{boyd1988culture}. The need to model the minds of others in common knowledge decision-making suggests that network properties should still be sufficient to separate these processes from rules that observe individuals, but do not model their internal states.

Our results here have used the language of risky tasks and fad adoption. Common knowledge is more than this, however. Implicated in everything from wage negotiation to land reform, it is a source of both stability~\cite{Ostrom01101994, aoki2007endogenizing} and change~\cite{change2008} in a society's institutions. In periods of upheaval, when conventions are rethought or rejected, common knowledge is damaged and individuals lose the ability to rely on their knowledge of what their neighbors will believe and accept. A historical example is provided by the chaotic state of the financial system during the French revolution at the end of the 18th Century. The emergence of multiple forms of credit-notes meant that individuals were forced to decide what forms of payment to accept based on what they expected others would later accept from them~\cite{spang2015stuff}.

Common knowledge requires that people communicate not just facts, but mutual knowledge of their states of mind. The difficulty of establishing common knowledge thus extends to electronic networks, even though communication costs there are low. The non-state currency BitCoin has drawn the attention of banking institutions, but it is only secondarily a source of value. Its proof-of-work protocol is a mechanism of achieving massively decentralized common knowledge about the contents of a ledger~\cite{satoshi}. In the case of Wikipedia, the need for common knowledge of norms leads to elaborate systems of norm discussion and referencing~\cite{heaberlin} and, even more literally, pages such as ``Assume the assumption of good faith''\footnote{See \url{https://en.wikipedia.org/wiki/Wikipedia:Assume_the_assumption_of_good_faith}} that explicitly prescribe beliefs users should hold about the beliefs of others. The era of electronic communication has only diversified the nature of the common knowledge problem. 

\section{Conclusion}

The need for common knowledge extends well beyond the problem of cooperative hunting, to include a vast range of endeavors, from joining a risky start-up to protesting an unjust regime. For cooperation in risky or complex tasks to be successful, participants must not only be motivated to undertake the task, but must also have the capability to successfully coordinate their efforts. 

Indeed, much of human culture is not simply what we do, but what we do given the mutual expectations in our social worlds. Common knowledge problems lie beyond simple contagion, and are some of the clearest examples of how culture is more than just the spread of memes adopted through observation alone.

Our results show that decision rules sensitive to the need for common knowledge leave clear traces on the behavior of individuals in a social network. Our results also show that common knowledge provides a new mechanism for the generation of inequality, over and above models that rely upon multiplicative accumulation of wealth.

\section*{Acknowledgments}

\noindent We thank Mirta Galesic, Colin Allen, and David Reed for helpful comments on this manuscript, and the Santa Fe Institute for hospitality while this work was completed. We thank John K. Kruschke for his assistance with Bayesian analysis and modeling, and Paul L. Hooper for discussions on anthropological research.

\clearpage

\begin{figure}
\begin{center}
\begin{tabular}{cc}
\includegraphics[scale=0.8]{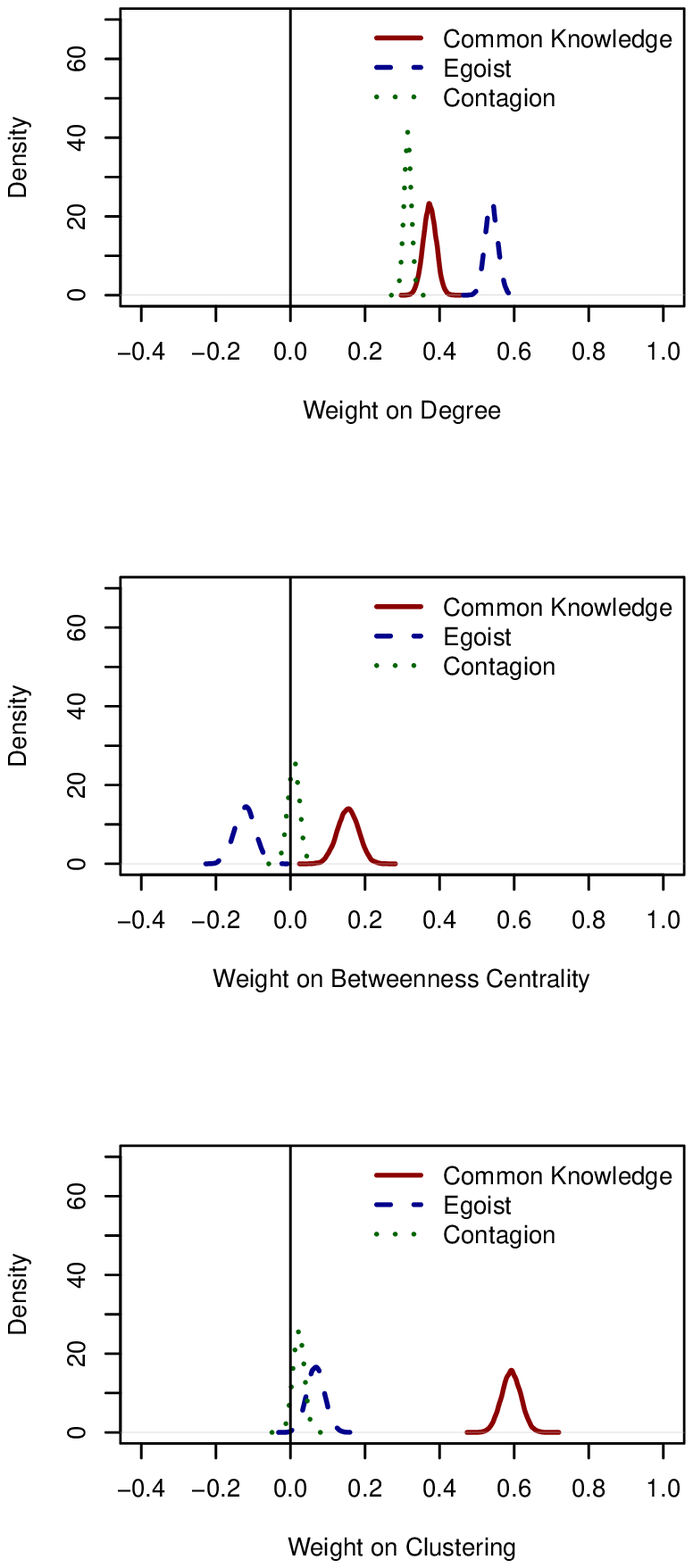}
\includegraphics[scale=0.8]{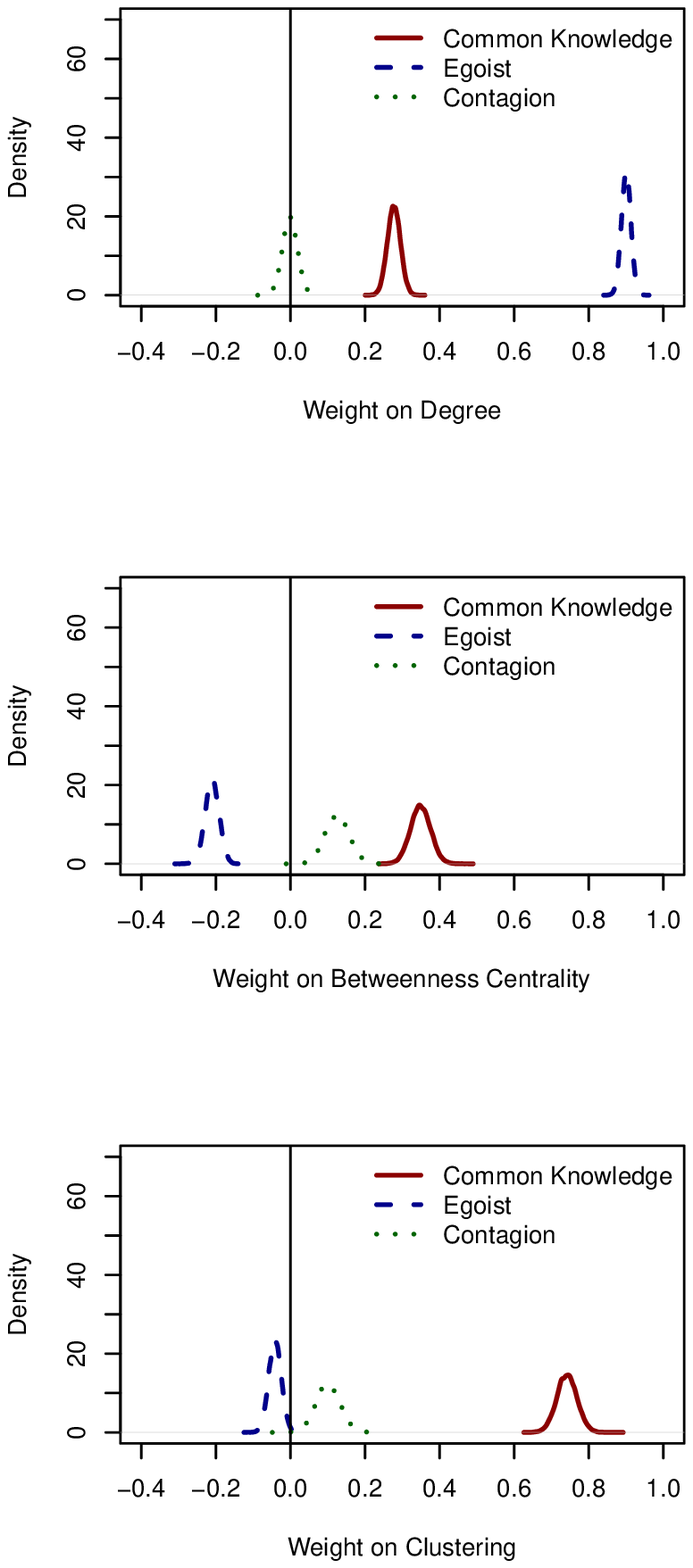}
\end{tabular}
\end{center}
\caption{\label{props} Posterior distributions on the regression weights for each of the three models in the hunting party scenario (left column) and the adopted fad scenario (right column). Clustering and (to a lesser extent) betweenness centrality are strong and unique predictors of task participation in the common knowledge case, but not in the less-sophisticated contagion and egoist models.}
\end{figure}

\appendix
\section*{Appendix}

\noindent In this Appendix, we present details of our simulation, and the parameter choices and justifications that form the basis of our results above. 

We also present data on ``quorum success'', a measure related to ``participation success'' used to quantify the results presented in the main text. As noted there, and as we show in detail here, the outcomes associated with quorum success are consistent with participation success.

\section{Simulation procedure}

For participation success, our outcome of interest is participation in a behavior. In the following simulations, we operationalize success as the number of timesteps the node participates in the behavior over the period of the simulation. A single simulation begins with a randomly generated graph, generated as described in Methods, Sec.~\ref{random_graph}, above. Following this, each of the three models is run on this graph structure for 200 rounds of potential cooperation. 

We model the decreasing enthusiasm described above as follows: every time an individual attends a hunt, their threshold is increased by one. If an individual does not attend a hunt and their threshold is above the base threshold, she then decreases her threshold by one. This process stops once the base threshold is reached. 

For the hunting party case, we choose parameters such that, for each decision rule, the success rates have means close to 80: $\theta=7$ and $\alpha=0.01$ for the common knowledge model, $\theta=6$ and $\alpha=0.01$ for the egoist model, and $\theta=5$ and $\alpha=0.05$ for the contagion model. For the adopted fad case, where participation is easier, we choose parameters such that the success rates have means close to 150: $\theta=8$ and $\alpha=0.01$ for the common knowledge model, $\theta=18$ and $\alpha=0.01$ for the egoist model and $\theta=4$ and $\alpha=0.05$ for the contagion model. 

\section{Details on Inequality Calculations}

\subsection{Hunting Party}

At each investigated parameter set, we generated 20 graphs using the generation method described above, and applied all three models to each graph.  We fixed the parameters of each model such that each model had an average success rate of approximately 80 rounds participating out of the 200. For the common knowledge model this meant $\theta=7$ and $\alpha=0.01$, for the egoist model this meant $\theta=6$ and $\alpha=0.01$, and for the contagion model this meant $\theta=5$ and $\alpha=0.05$. The number of rounds spent participating was recorded for each node. Then the distribution of success across this sample of networks was plotted as a histogram, as shown in Figure~\ref{fig:Hists}. We found that the distribution of success was more unequal for the egoist and common knowledge models, with clusters of high success and low success individuals, whereas the contagion model demonstrated comparative equality, with a single unimodal cluster. To quantitatively assess the inequality encouraged by the two models, we computed the Gini coefficient~\citep{Gini1912} of the distribution of success for each model. The Gini coefficient is a commonly used measure of inequality~\citep{Deininger1996,Thomas2001,Yitzhaki1979} that measures the deviation of a distribution (in our case, the distribution of success) from perfect equality, on a scale from 0 to 1. A Gini coefficient of 0 indicates perfect equality, with all individuals having equal values of success, and a Gini coefficient of 1 indicates total inequality, with a single individual having high levels of success and all others have none. The Gini coefficient for the common knowledge distribution is 0.159, the coefficient for the egoist distribution is 0.221, and the Gini coefficient for the contagion model is 0.055. We also computed the Gini coefficients for each of the 20 generated networks individually, instead of collapsed across all the generated networks, as presented here. For all networks, the Gini coefficient is smaller for the contagion model than for either the common knowledge model or the egoist model. This supports the qualitative judgment that inequality is encouraged by the common knowledge and egoist models, and not by the contagion model. 

\subsection{Adopted Fad}

We again investigated the distribution of success across nodes in each of the three models. In this scenario, we chose parameter values such that each model had an average success rate of approximately 150 rounds participating out of 200. Participation lasts for multiple rounds at a time in this scenario, thus 150 rounds out of 200 is comparable to the 80 rounds of success utilized in the hunt scenario. For the common knowledge model, we set $\theta=8$ and $\alpha=0.01$, for the egoist model we set $\theta=18$ and $\alpha=0.01$, and for the contagion model we set $\theta=4$ and $\alpha=0.05$. Visual inspection indicated a pattern similar to the previous scenario: under these parameter values the egoist and common knowledge models produce unequal and multimodal distributions of success, whereas the contagion model produces relative equality and a unimodal distribution. Analysis of the Gini coefficient confirms this: the common knowledge distribution has a Gini coefficient of 0.147, the egoist distributions has a Gini coefficient of 0.158, and the contagion distribution has a Gini coefficient of 0.029.

We again computed the Gini coefficient for each network individually, and again found that the contagion Gini coefficient never exceeded the Gini coefficient for either of the other two models. This again quantitatively confirms that the common knowledge and egoist models demonstrate higher levels of inequality than the contagion model.

\section{Details on Node Property Calculations}

\subsection{Hunting Party}

We generated 50 random graphs. Then, for each graph, we applied all three models and recorded information about each node's success, as well as a set of node properties (described below). We used the same parameter values described in the previous section for the three models. However, it should be noted that the results reported here are robust to changing parameter settings: in all parameter settings we tested where meaningful regression was not ruled out (graphs with no variation in number of successful hunts) the same qualitative pattern of regression weights was observed.

To analyze the relationship between success and node properties we used a hierarchical multiple linear regression, with the parameters of the model estimated via Bayesian methods. 

The weight on degree is credibly non-zero and positive for the common knowledge (Mode = 0.372, 95\% HDI from 0.339 to 0.407), egoist (Mode = 0.541, 95\% HDI from 0.508 to 0.573), and contagion (Mode = 0.315, 95\% HDI from 0.295 to 0.334) models.

To assess whether a parameter value is credibly non-zero, we compare the 95\% highest density interval (HDI) of the posterior distribution to the null value. The 95\% HDI is analogous to a confidence interval, but represents the bulk of the distribution such that no parameter value outside the HDI has greater credibility than parameter values inside the HDI. The null value is represented by a region of practical equivalence (ROPE) around the null value. For all comparisons presented here, we used a ROPE of -0.1 to 0.1 on the scale of the standardized regression coefficients. If the HDI falls outside this range, we conclude that the parameter is credibly non-zero. If the HDI falls entirely with this range, we conclude that the parameter value is practically equivalent to zero.

The effect of betweenness centrality differs slightly across the three models: it is a marginally positive predictor of success in the common knowledge model (Mode = 0.156, 95\% HDI from 0.098 to 0.208), a marginally negative predictor in the egoist model (Mode = -0.119, 95\% HDI from -0.176 to -0.067), and is practically equivalent to zero in the contagion model (Mode = 0.012, 95\% HDI from -0.022 to 0.039).  In the case of clustering, the differences across the model are stark. The weight is credibly positive in the common knowledge model (Mode = 0.592, 95\% HDI from 0.542 to 0.643), positive but not credibly non-zero in the egoist model (Mode = 0.069, 95\% HDI from 0.021 to 0.115) and practically equivalent to zero in the contagion model (Mode = 0.024, 95\% HDI from -0.008 to 0.051). The strong weight on clustering is unique to the common knowledge model. This is the primary signature that common knowledge is operating: a common knowledge decision rule predicts that those with densely connected local neighborhoods will be the most successful.

\subsection{Adopted Fad}

In this scenario, the weight on degree is credibly non-zero for the common knowledge (Mode = 0.275, 95\% HDI from 0.243 to 0.312) and egoist (Mode = 0.900, 95\% HDI from 0.874 to 0.925) models, but is practically equivalent to zero for the contagion model (Mode = -0.003, 95\% HDI from -0.041 to 0.038). Moreover, the effect of degree in the egoist model is much greater than the weight in the other two models. The effect of betweenness centrality differs across the three models: it is a credibly positive predictor of success in the common knowledge model (Mode = 0.346, 95\% HDI from 0.295 to 0.401), a credibly negative predictor in the egoist model (Mode = -0.212, 95\% HDI from -0.246 to -0.173), and is marginally positive in the contagion model (Mode = 0.126, 95\% HDI from 0.064 to 0.188). The case of clustering is very similar to the hunt scenario. The weight in contagion (Mode = 0.097, 95\% HDI from 0.038 to 0.165) is marginally positive, the weight in the egoist model (Mode = -0.042, 95\% HDI from -0.074 to -0.007) is practically equivalent to zero, and the weight in the common knowledge is credibly positive (Mode = 0.745, 95\% HDI from 0.688 to 0.795). This again confirms that a major signature that common knowledge is operating is that those with densely connected local neighborhoods will be most successful.

\section{Quorum Success}
In the analyses presented so far, our operationalization of success was participation. Another potential operationalization of success is a combination of participation \emph{and} the presence of enough participating neighbors to meet the node's threshold. We call this definition ``quorum success''. Under the quorum definition, participation is determined by the decision rules described above for each model, and following this participation decision, the number of participating neighbors is counted and the node is successful only if this count meets the node's threshold. Non-attempting rounds are not counted as successes regardless of the number of participating neighbors, just as before.

To ensure that the results presented here were generalizable, we implemented this rule and repeated the analyses presented above on the success behavior that resulted.

\subsection{Inequality}
\begin{figure}
\begin{center}
\includegraphics[width=\textwidth]{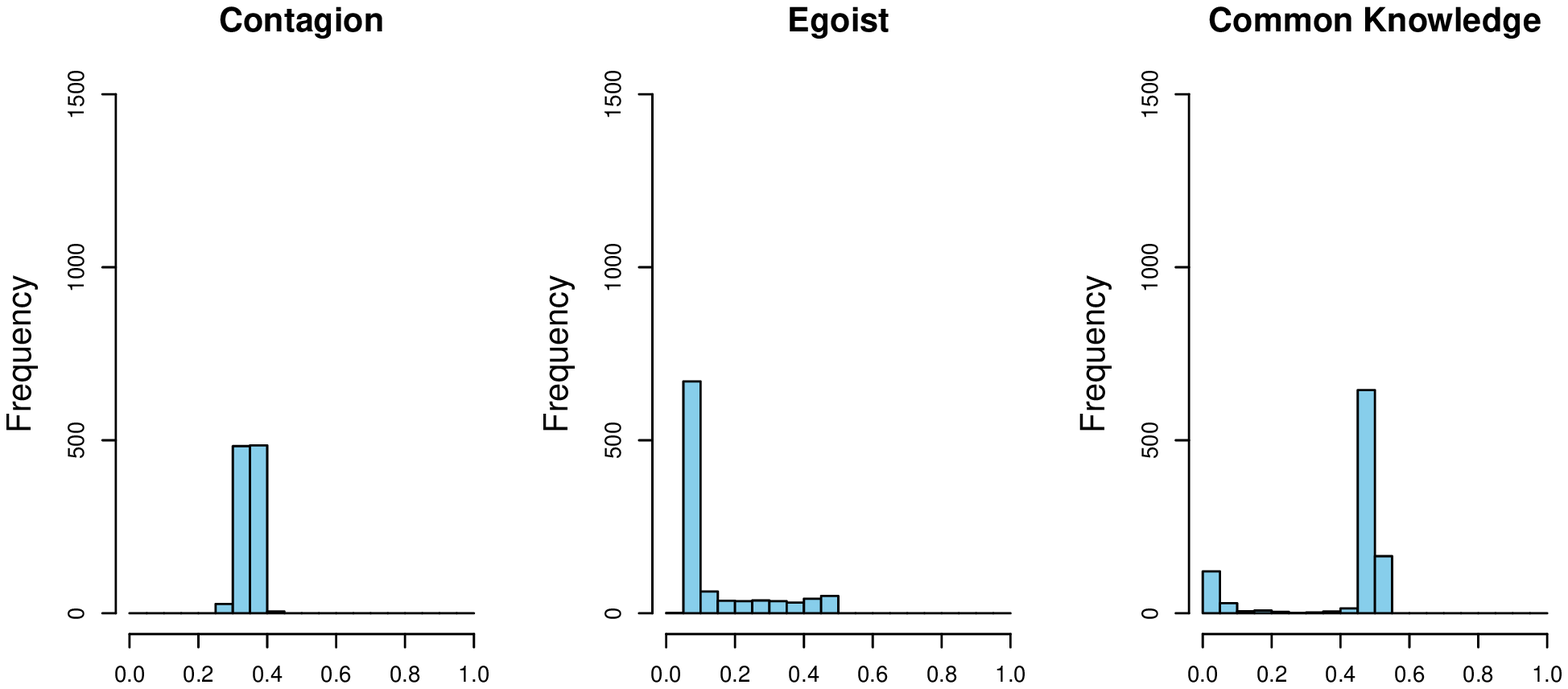} \\
\includegraphics[width=\textwidth]{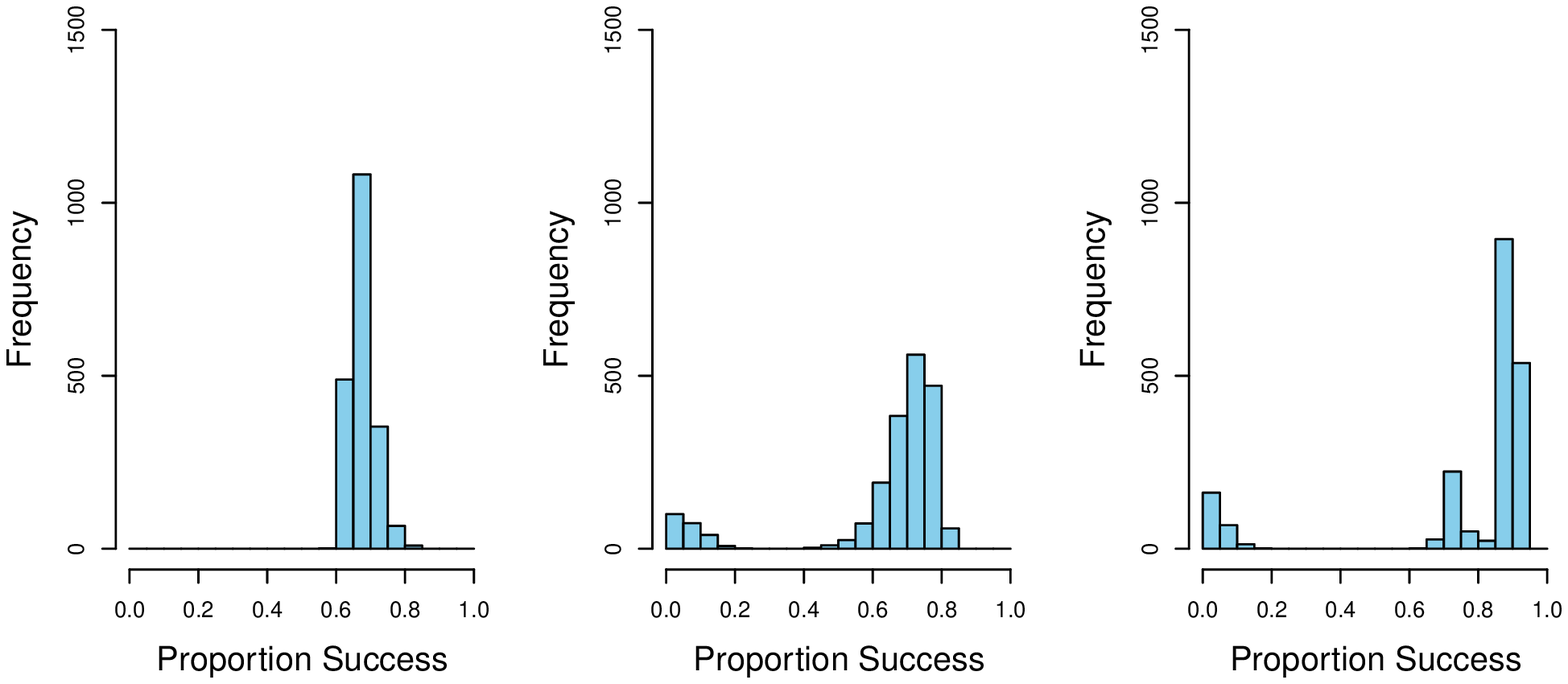}
\end{center}
\caption{\label{fig:SuccessHists} Distributions of individual participation rates for the hunting party scenario (top row) and fad adoption (bottom row) using the operationalization of success described in this section. Notice that the patterns are very similar to Figure~\ref{fig:Hists} with the exception of the Egoist distribution in the hunting party scenario. Success in this circumstance no longer has a large group of individuals who all have high levels of success. This is because the egoist decision rule in the hunt scenario often encourages participation when a node's neighbors are not guaranteed to participate.}
\end{figure}

Our analysis here uses the same procedure described previously, with the same parameter values for each of the three models. Figure~\ref{fig:SuccessHists} shows the distribution of success under the quorum rule, and Table~\ref{SuccessGini} reports the Gini coefficients of the success distributions for each of the models in both scenarios. The results are very similar to the original success definition with one exception: the egoist model in the hunting party scenario. In this scenario, many times an egoist node will see neighbors that seem to them to be willing to participate, but in fact do not have sufficient neighbors themselves to participate. This means that participation often occurs in scenarios where not enough neighbors participate to meet a node's threshold. The result of this is that many nodes that were highly successful under the participation success rule are now unsuccessful in the quorum success rule. This is reflected in the lack of a second high mode in the distribution of success, and a very high Gini coefficient, as a much smaller proportion of the individuals have high success.

\begin{table}
\begin{tabular}{l|l|l}
Scenario & Decision Rule & Gini Coefficient \\ \hline
Hunting Party & Contagion & 0.047 \\
& Egoist & 0.420 \\
& Common Knowledge & 0.163 \\ \hline
Adopted Fad & Contagion & 0.028 \\
& Egoist & 0.145 \\
& Common Knowledge & 0.144
\end{tabular}
\caption{\label{SuccessGini} Inequalities of outcome: In the alternative success definition, the egoist and common knowledge still rules produce far more unequal outcomes (higher Gini coefficients) than the more ``egalitarian'' contagion process. As in the histograms, the egoist decision rule in the hunting party scenario is the only major area of difference: due to the lack of the large group of high performers, the level of inequality is higher under this success rule than the participation success rule.}
\end{table}

\subsection{Node Properties}
\begin{figure}
\begin{center}
\begin{tabular}{cc}
\includegraphics[scale=0.8]{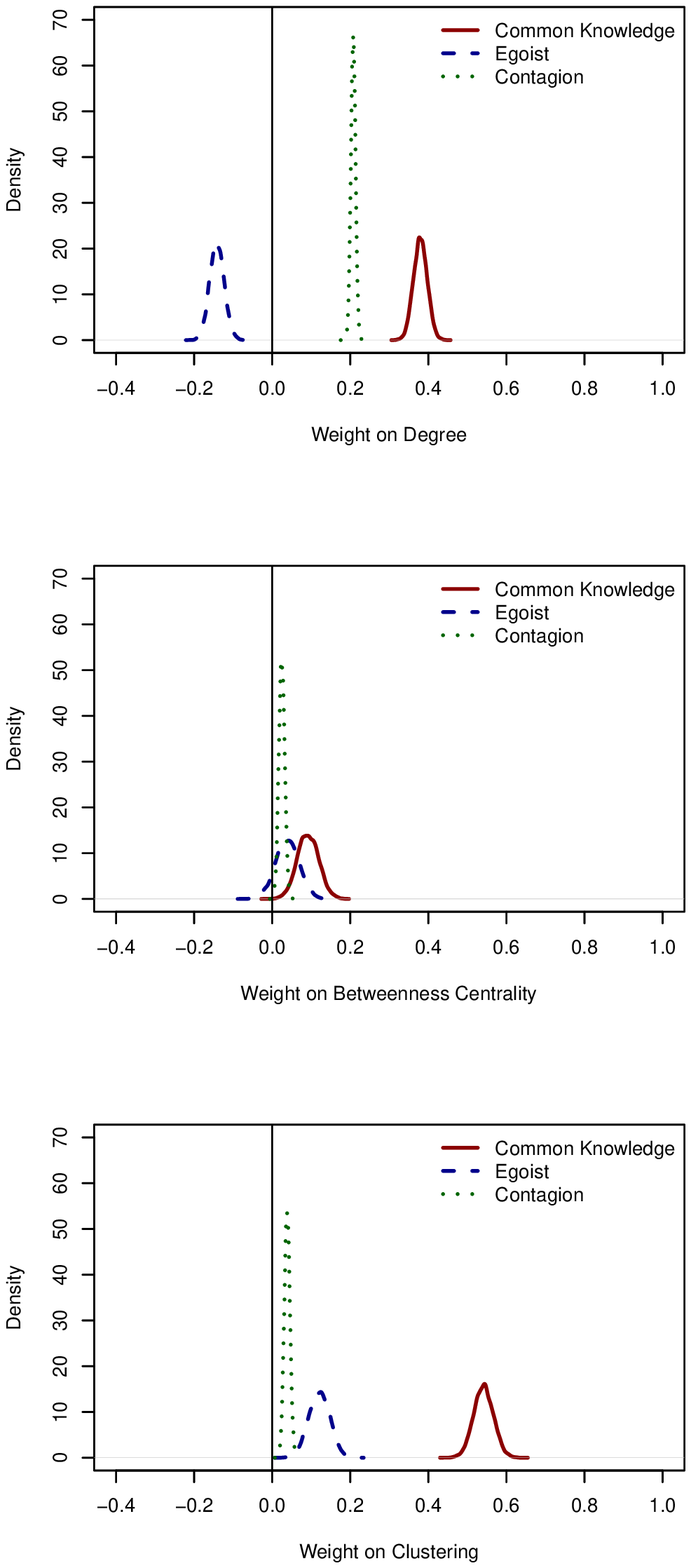}
\includegraphics[scale=0.8]{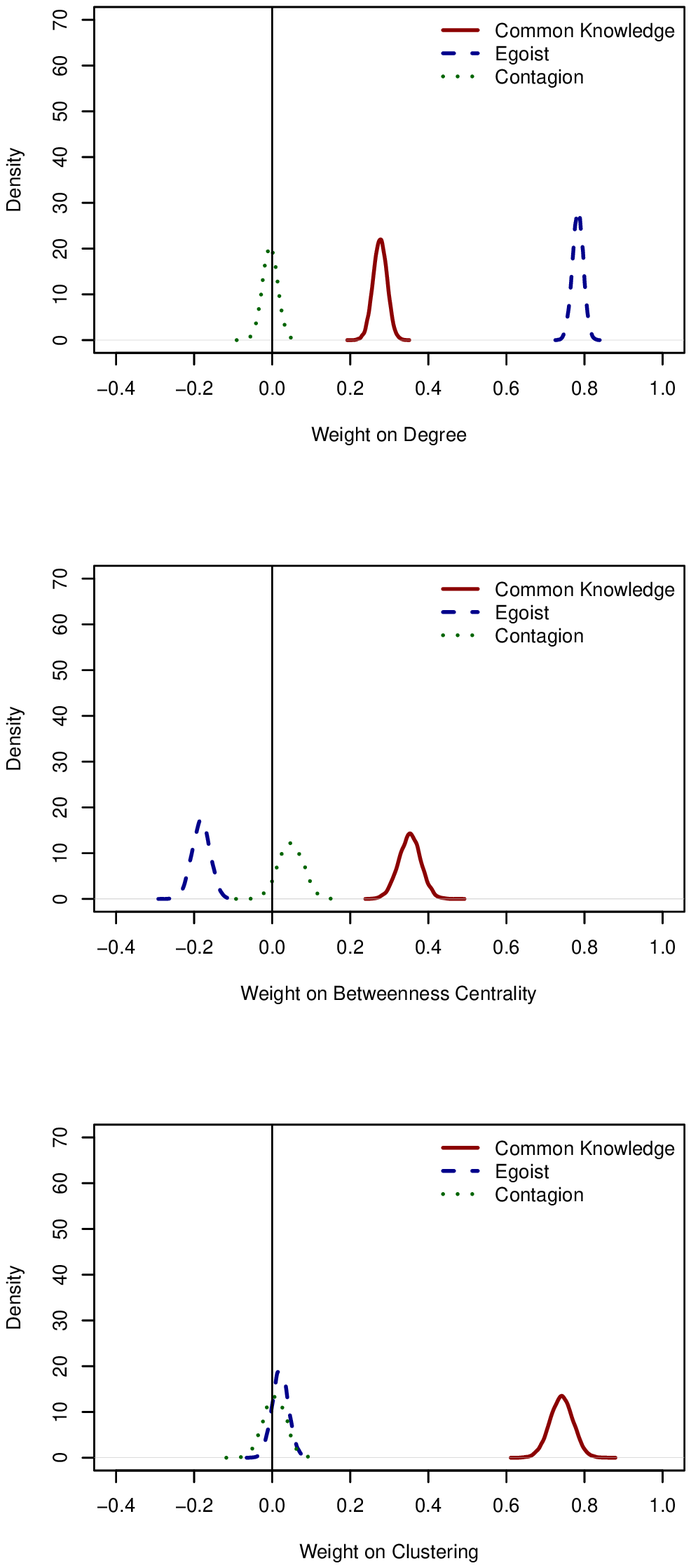}
\end{tabular}
\end{center}
\caption{\label{succProps} Posterior distributions on the regression weights for each of the three models in the hunting party scenario (left column) and the adopted fad scenario (right column) using the quorum success rule. The adopted fad results are nearly the same as the results using the participation success rule. The results from the hunt scenario are very similar to the results using the participation success rule, with the exception of the weight on degree in the egoist model, which is now negative.}
\end{figure}

We also analyzed the properties that make a node successful using the quorum definition of success, again using parameters and analysis methods identical to those previously described. Figure~\ref{succProps} shows the posterior distributions on the regression weights in both scenarios using the quorum success rule. As in the inequality analysis, the results in the fad scenario are largely identical to the participation success rule; all credibly non-zero effects remain credibly non-zero, and no new credibly non-zero effects are observed. The results in the hunt case also closely mirror the inequality results: the egoist model is unique in that it has importantly different results when using the quorum success rule than when using the participation success rule. The egoist model has a credibly negative effect of degree (where before it was positive) and a practically equivalent to zero effect of betweenness (where before it was marginally negative). This negative effect of degree is at first puzzling: despite degree being a strong positive predictor of participation (and participation being a necessary condition for success) it is actually a negative predictor of success. We believe this effect results from those with high degree being at higher risk of becoming ``out of sync'' with their neighbors. Recall that in the hunt scenario, individuals are less likely to hunt shortly after they have already hunted. If a node has a small but sufficient number of neighbors that tend to successfully hunt together, adding additional neighbors could cause the node to participate as a result of these new neighbors when the smaller (but successful) hunting group is still in an unwilling to participate state. If this larger group is not synced, this will lead to failure. Following this, the node will itself be unwilling to participate, and the small successful group will attend without it.

\subsubsection{Quantitative Details: Hunt Scenario}

Here we present quorum success results for the hunt scenario.. The weight on degree is credibly non-zero and positive for the common knowledge (Mode = 0.377, 95\% HDI from 0.345 to 0.415), credibly negative for the egoist model (Mode = -0.144, 95\% HDI from -0.177 to -0.105), and credibly positive for contagion (Mode = 0.208, 95\% HDI from 0.196 to 0.220) models. The effect of betweenness centrality differs slightly across the three models: it is a marginally positive predictor of success in the common knowledge model (Mode = 0.090, 95\% HDI from 0.040 to 0.147), and a practically equivalent to zero effect in the egoist model (Mode = 0.042, 95\% HDI from -0.020 to 0.099) and the contagion model (Mode = 0.023, 95\% HDI from 0.010 to 0.039).  As before, the case of clustering has large differences across the models. The weight is credibly positive in the common knowledge model (Mode = 0.544, 95\% HDI from 0.494 to 0.593), positive but not credibly non-zero in the egoist model (Mode = 0.124, 95\% HDI from 0.068 to 0.172) and practically equivalent to zero in the contagion model (Mode = 0.038, 95\% HDI from 0.024 to 0.053). Once again, the strong weight on clustering is unique to the common knowledge model. 

\subsubsection{Quantitative Details: Adopted Fad}

Here we present quorum success results for the adopted fad scenario. In this case, the weight on degree is credibly non-zero for the common knowledge (Mode = 0.278, 95\% HDI from 0.241 to 0.311) and egoist (Mode = 0.783, 95\% HDI from 0.754 to 0.811) models, but is practically equivalent to zero for the contagion model (Mode = -0.004, 95\% HDI from -0.042 to 0.034). The effect of betweenness centrality again differs across the three models: it is a credibly positive predictor of success in the common knowledge model (Mode = 0.353, 95\% HDI from 0.298 to 0.410), a credibly negative predictor in the egoist model (Mode = -0.182, 95\% HDI from -0.227 to -0.137), and is very slightly positive in the contagion model (Mode = 0.044, 95\% HDI from -0.017 to 0.106). Finally, for clustering the weight in the common knowledge model is credibly positive (Mode = 0.742, 95\% HDI from 0.684 to 0.802), the weight in the egoist model (Mode = 0.015, 95\% HDI from -0.021 to 0.062) is practically equivalent to zero, and the weight in contagion (Mode = 0.001, 95\% HDI from -0.049 to 0.062) is practically equivalent to zero. This again confirms that a major signature that common knowledge is operating is that those with densely connected local neighborhoods will be most successful.

\clearpage
\bibliographystyle{unsrt}
\bibliography{CKonNetworks}

\begin{thebibliography}{10}

\bibitem{Skyrms2001}
Brian Skyrms.
\newblock {The Stag Hunt}.
\newblock {\em Proceedings and Addresses of the American Philosophical
  Association}, 75(2):31--41, 2001.

\bibitem{Chwe1999}
Michael Suk-Young Chwe.
\newblock {Structure and strategy in collective action}.
\newblock {\em American Journal of Sociology}, 105(1):128--156, 1999.

\bibitem{Bowles2011}
Samuel Bowles and Herbert Gintis.
\newblock {\em {A Cooperative Species: Human Reciprocity and its Evolution}}.
\newblock Princeton University Press, Princeton, New Jersey, 2011.

\bibitem{Gigone1993}
Daniel Gigone and Reid Hastie.
\newblock The common knowledge effect: Information sharing and group judgment.
\newblock {\em Journal of Personality and Social Psychology}, 65(5):959, 1993.

\bibitem{Stasser1992}
Garold Stasser and Dennis Stewart.
\newblock Discovery of hidden profiles by decision-making groups: Solving a
  problem versus making a judgment.
\newblock {\em Journal of personality and social psychology}, 63(3):426, 1992.

\bibitem{Reimer2010}
Torsten Reimer, Andrea Reimer, and Verlin~B Hinsz.
\newblock Na{\"\i}ve groups can solve the hidden-profile problem.
\newblock {\em Human Communication Research}, 36(3):443--467, 2010.

\bibitem{Aumann1995}
Robert~J Aumann and Adam Brandenburger.
\newblock {Epistemic conditions for Nash equilibrium}.
\newblock {\em Econometrica}, 63(5):1161--1180, 1995.

\bibitem{Aumann1999}
Robert~J Aumann.
\newblock {Interactive epistemology I: Knowledge}.
\newblock {\em International Journal of Game Theory}, 28(3):263--300, August
  1999.

\bibitem{john1992}
John Geanakoplos.
\newblock Common knowledge.
\newblock {\em Journal of Economic Perspectives}, 6(4):53--82, 1992.

\bibitem{Gintis2009}
Herbert Gintis.
\newblock {\em {The bounds of reason: Game theory and the unification of the
  behavioral sciences}}.
\newblock Princeton University Press, Princeton, New Jersey, September 2009.

\bibitem{Chwe2013}
Michael Suk-Young Chwe.
\newblock {\em {Rational ritual: Culture, coordination, and common knowledge}}.
\newblock Princeton University Press, Princeton, New Jersey, 2013.

\bibitem{Camerer2004}
Colin~F Camerer, Teck-Hua Ho, and Juin-Kuan Chong.
\newblock {A cognitive hierarchy model of games}.
\newblock {\em The Quarterly Journal of Economics}, 119(August):861--898, 2004.

\bibitem{Lee2012}
Ritchie Lee and David Wolpert.
\newblock {Game theoretic modeling of pilot behavior during mid-air
  encounters}.
\newblock In {\em Decision Making with Imperfect Decision Makers}, pages
  75----111. Springer, 2012.

\bibitem{Granovetter1978}
Mark~S. Granovetter.
\newblock {Threshold Models of Collective Behavior}.
\newblock {\em The American Journal of Sociology}, 83(6):1420--1443, 1978.

\bibitem{Watts2002}
Duncan~J Watts.
\newblock {A simple model of global cascades on random networks.}
\newblock {\em Proceedings of the National Academy of Sciences of the United
  States of America}, 99(9):5766--5771, 2002.

\bibitem{ross1977false}
Lee Ross, David Greene, and Pamela House.
\newblock The “false consensus effect”: An egocentric bias in social
  perception and attribution processes.
\newblock {\em Journal of Experimental Social Psychology}, 13(3):279--301,
  1977.

\bibitem{krueger1994truly}
Joachim Krueger and Russell~W Clement.
\newblock The truly false consensus effect: an ineradicable and egocentric bias
  in social perception.
\newblock {\em Journal of Personality and Social Psychology}, 67(4):596, 1994.

\bibitem{Dawes1989}
Robyn~M. Dawes.
\newblock Statistical criteria for establishing a truly false consensus effect.
\newblock {\em Journal of Experimental Social Psychology}, 25(1):1--17, 1989.

\bibitem{Watts1998}
Duncan~J. Watts and Steven~H. Strogatz.
\newblock {Collective dynamics of small-world networks}.
\newblock {\em Nature}, 393:440--442, 1998.

\bibitem{Hagberg2008}
Aric~A. Hagberg, Daniel~A. Schult, and Pieter~J. Swart.
\newblock {Exploring network structure, dynamics, and function using NetworkX}.
\newblock {\em Proceedings of the 7th Python in Science Conference}, 2008.

\bibitem{Gini1912}
CW~Gini.
\newblock Variability and mutability, contribution to the study of statistical
  distribution and relaitons.
\newblock {\em Studi Economico-Giuricici della R}, 1912.

\bibitem{Kruschke2015}
John~K Kruschke.
\newblock {\em {Doing Bayesian data analysis, second edition: A tutorial with
  R, JAGS, and Stan}}.
\newblock Academic Press/Elsevier, Burlington, MA, 2nd edition, 2015.

\bibitem{Kruschke2015b}
John~K Kruschke and Torrin~M Liddell.
\newblock Bayesian data analysis for newcomers.
\newblock 2015.
\newblock In prep.

\bibitem{piketty2001income}
Thomas Piketty and Emmanuel Saez.
\newblock Income inequality in the united states, 1913--1998.
\newblock Technical report, National Bureau of Economic Research, 2001.

\bibitem{Piketty01082014}
Thomas Piketty and Gabriel Zucman.
\newblock Capital is back: Wealth-income ratios in rich countries 1700--2010.
\newblock {\em The Quarterly Journal of Economics}, 129(3):1255--1310, 2014.

\bibitem{duncan2005apple}
Greg Duncan, Ariel Kalil, Susan~E Mayer, Robin Tepper, and Monique~R Payne.
\newblock The apple does not fall far from the tree.
\newblock In Samuel Bowles, Herbert Gintis, and Melissa~Osborne Groves,
  editors, {\em Unequal chances: Family background and economic success}, pages
  23--79. Russell Sage Foundation, 2005.

\bibitem{e15114932}
Paul~L. Hooper, Simon DeDeo, Ann~E. Caldwell~Hooper, Michael Gurven, and
  Hillard~S. Kaplan.
\newblock Dynamical structure of a traditional amazonian social network.
\newblock {\em Entropy}, 15(11):4932, 2013.

\bibitem{boyd1988culture}
Robert Boyd and Peter~J Richerson.
\newblock {\em Culture and the evolutionary process}.
\newblock University of Chicago Press, 1988.

\bibitem{Ostrom01101994}
Elinor Ostrom.
\newblock Constituting social capital and collective action.
\newblock {\em Journal of Theoretical Politics}, 6(4):527--562, 1994.

\bibitem{aoki2007endogenizing}
Masahiko Aoki.
\newblock Endogenizing institutions and institutional changes.
\newblock {\em Journal of Institutional Economics}, 3(01):1--31, 2007.

\bibitem{change2008}
Pepper~D. Culpepper.
\newblock The politics of common knowledge: Ideas and institutional change in
  wage bargaining.
\newblock {\em International Organization}, 62(1):pp. 1--33, 2008.

\bibitem{spang2015stuff}
Rebecca~L. Spang.
\newblock {\em Stuff and Money in the Time of the French Revolution}.
\newblock Harvard University Press, 2015.

\bibitem{satoshi}
Satoshi Nakamoto.
\newblock Re: Bitcoin p2p e-cash paper, 2008.
\newblock The Cryptography and Cryptography Policy Mailing List. 13 November
  2008. 19:04:25 -0800. msg09997.
  \url{http://www.mail-archive.com/cryptography@metzdowd.com/msg09997.html}.

\bibitem{heaberlin}
Bradi Heaberlin and Simon DeDeo.
\newblock {Norm Bundles on Wikipedia}.
\newblock 2015.
\newblock In prep.

\bibitem{Deininger1996}
Klaus Deininger and Lyn Squire.
\newblock A new data set measuring income inequality.
\newblock {\em The World Bank Economic Review}, 10(3):565--591, 1996.

\bibitem{Thomas2001}
Vinod Thomas, Yan Wang, and Xibo Fan.
\newblock {\em {Measuring education inequality: Gini coefficients of
  education}}, volume 2525.
\newblock World Bank Publications, 2001.

\bibitem{Yitzhaki1979}
Shlomo Yitzhaki.
\newblock {Relative deprivation and the Gini coefficient}.
\newblock {\em The Quarterly Journal of Economics}, pages 321--324, 1979.

\end{thebibliography}
\end{document}